\documentstyle[12pt,epsf,epsfig,cite]{article}

\newcommand{\beq}{\begin{equation}}
\newcommand{\eeq}{\end{equation}}
\newcommand{\bea}{\begin{eqnarray}}
\newcommand{\eea}{\end{eqnarray}}

\newcommand{\gsim}{\lower.7ex\hbox{$
\;\stackrel{\textstyle>}{\sim}\;$}}
\newcommand{\lsim}{\lower.7ex\hbox{$
\;\stackrel{\textstyle<}{\sim}\;$}}

\def\lsim{\mathrel{\rlap{\lower3pt\hbox{\hskip0pt$\sim$}}
    \raise1pt\hbox{$<$}}}         
\def\gsim{\mathrel{\rlap{\lower4pt\hbox{\hskip1pt$\sim$}}
    \raise1pt\hbox{$>$}}}         

\begin{document}

\begin{titlepage}
\pagestyle{empty}
\baselineskip=21pt
\rightline{hep-ph/0211461}
\rightline{ }
\vskip 0.05in
\begin{center}
{\Large{\bf Charting the Higgs Boson Profile\\ at $e^+e^-$ Linear Colliders}}
\end{center}
\begin{center}
\vskip 0.05in
{{\bf Marco~Battaglia}}
\vskip 0.05in
{\it {CERN, Geneva, Switzerland}
}
\vskip 0.15in
{\bf Abstract}
\end{center}
\baselineskip=18pt \noindent
The problems of the origin of mass and of electro-weak symmetry breaking are 
central to the programme of research in particle physics, at present and in 
the coming decades. This paper reviews the potential of high energy, high
luminosity $e^+e^-$ linear colliders in exploring the Higgs sector,  
to extend and complement the data which will become available from hadron 
colliders. The accuracy of measurements of the Higgs boson properties
will not only probe the validity of the Higgs mechanism but also provide 
sensitivity to New Physics beyond the Standard Model.

\vspace{0.35cm}

\begin{center}
\epsfig{file=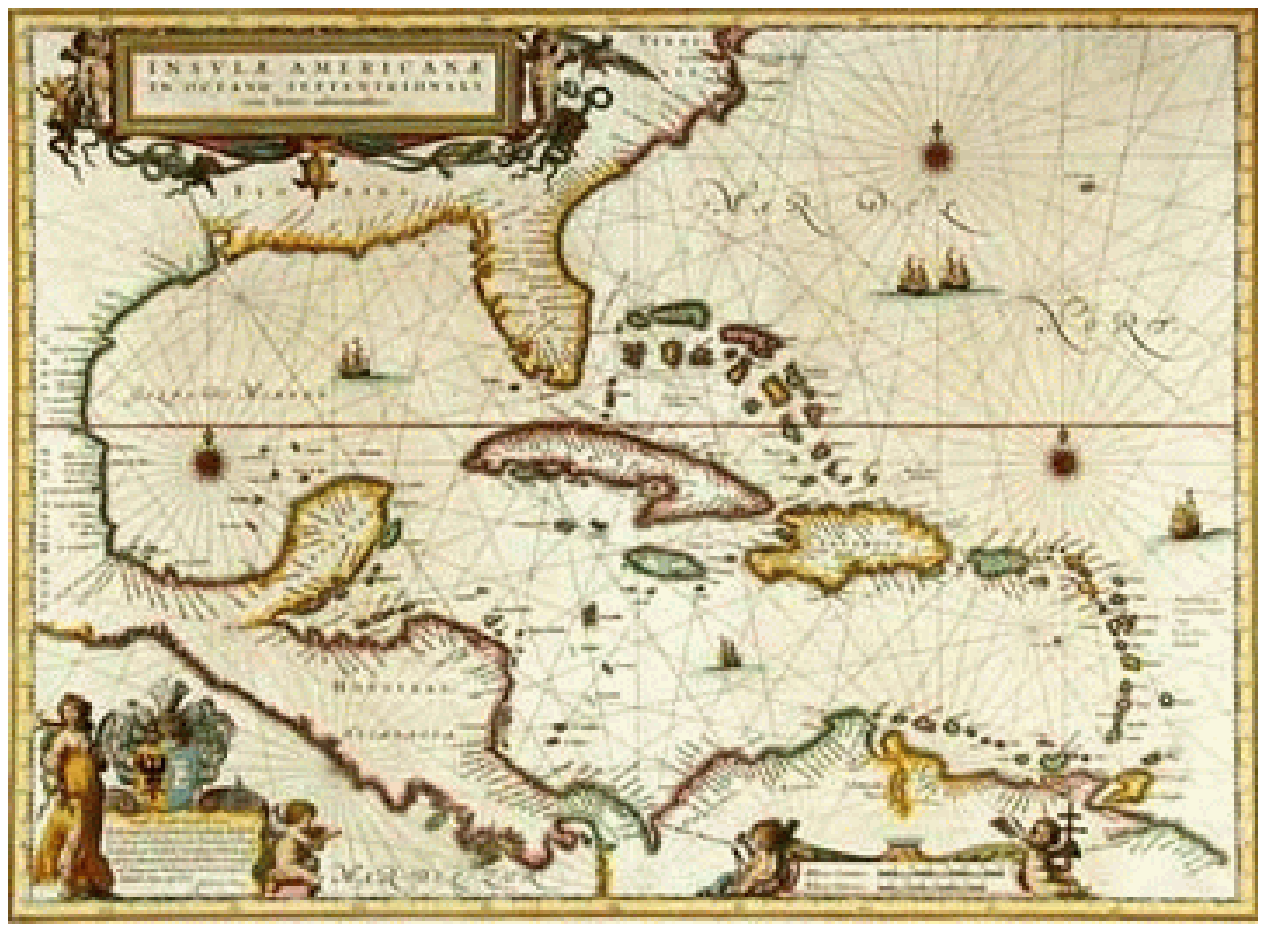,width=7.0cm}
\end{center}

\vfill
\begin{center}{\small Invited Talk at the 10$^{th}$ International Conference on Supersymmetry\\
and Unification of Fundamental Interactions\\
June 17-23, 2002, DESY, Hamburg}\end{center}

\vspace{0.75cm}

\begin{center}{\footnotesize \it This paper is dedicated to Laura Alidori}\end{center}
\vskip 0.15in
\leftline{  }
\leftline{2 November 2002}
\end{titlepage}
\baselineskip=18pt

\section{Prologue}

Explaining the origin of mass is one of the great scientific quests of our time. 
The Standard Model (SM), successfully tested to an unprecedented level of accuracy 
by the LEP, Tevatron  and SLC experiments, and now also by the $B$-factories, addresses 
this question with the Higgs mechanism~\cite{Higgs}. 

In this paper, I review the 
anticipated potential of $e^+e^-$ linear colliders (LC) in probing the nature of 
the Higgs sector. As the title suggests, such studies might be compared to the efforts 
of explorers and cartographers, from the first sightings of new lands to the 
systematic charting of their coastline profiles.  
The first manifestation of the Higgs mechanism through the Higgs sector is the 
existence of at least one Higgs boson, denoted with $H^0$. The observation of a new 
spin-0 particle would represent a first sign that the Higgs mechanism of mass 
generation is realised in Nature. Results of direct searches and indirect constraints, 
from precise electro-weak data, tell us that the Higgs boson is heavier than 
114~GeV and possibly lighter than about 195~GeV~\cite{ichep}. 
We expect the Higgs boson to be first sighted by the Tevatron or the {\sc Lhc}, 
the CERN hadron collider, which will determine its mass and perform a first survey of 
its basic 
properties. Similarly to the first sighting of Hispaniola in 1492, such a discovery 
will represent a major breakthrough and will bring to a successful completion an 
intense program of experimental searches and phenomenological speculations lasting since 
several decades. But after the observation of a new particle with properties compatible 
with those expected for the Higgs boson, a significant experimental and theoretical 
effort will be needed to verify that the newly-discovered particle is indeed
the boson of the scalar field responsible for the electro-weak symmetry breaking and 
the generation of mass. An $e^+e^-$ linear collider with its well defined energy, known 
identity of the initial state partons, opportunity to control their helicities and a 
detectors which provides highly accurate information on the event properties, will 
promote the Higgs studies into the domain of precision physics.

\section{$e^+e^-$ collisions from beyond LEP-2 to\\ the multi-TeV scale}

The {\sc Lep-2} collider at CERN has set the highest centre-of-mass energy, reached so 
far in $e^+e^-$ collisions, at $\sqrt{s}$=209~GeV. A significant further increase in 
both the beam energy and the luminosity has been the aim of several decades of LC designs
and R\&D, to advance the energy frontier in $e^+e^-$ physics. 
These developments have now matured to the point where we can 
contemplate construction of a linear collider with initial energy in the 500~GeV 
range and a credible upgrade path to $\sim1$~TeV. The {\sc Tesla} 
project~\cite{Brinkmann:2001qn} adopts super-conducting (SC) cavities which offer high 
luminosity with rather relaxed alignment requirements. As a result of a 
successful R\&D program, gradients in excess of 23~MV/m, necessary to reach 
$\sqrt{s}$=500~GeV have been demonstrated. An alternative approach is taken by the 
{\sc Nlc}~\cite{nlc} and {\sc Jlc} projects which adopt X-band (11~GHz) warm cavities, 
evolving the concept of the {\sc Slc} at SLAC, the only linear collider operated so far. 

\begin{figure}
\begin{center}
\epsfig{file=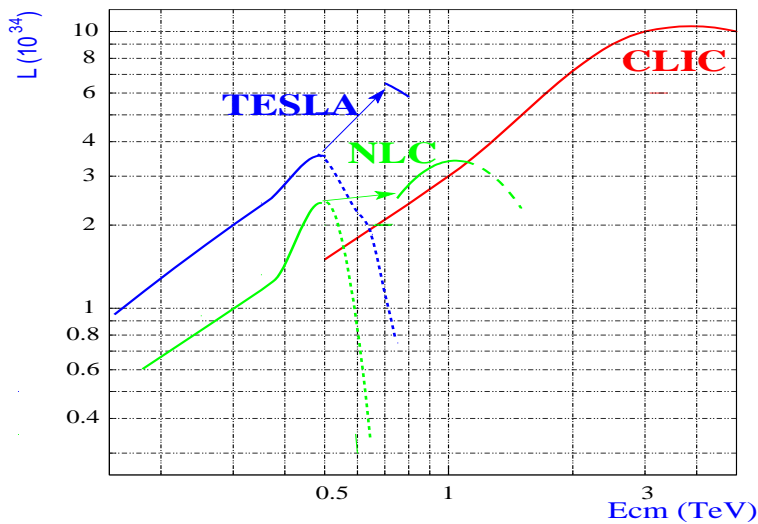,width=10.0cm,height=7.0cm,clip}
\end{center}
\vspace*{-0.5cm}
\caption[]{\sl The design luminosity as a function of the centre-of-mass energy, 
$\sqrt{s}$, for the {\sc Tesla} (SC cavities), {\sc Nlc} (X-band RF) and 
{\sc Clic} (two beam acceleration) projects. The linear collider has the potential 
to span its energy range over one order of magnitude. The dotted lines represent 
the performance for energy increases beyond 500~GeV with fixed power. The arrows show 
step in performance obtained by increasing the gradient 
(modified from~\cite{Battaglia:2002ns}).}
\label{fig:1}
\end{figure}

The LC energy can be later upgraded by three different strategies 
(see Figure~\ref{fig:1}). The first is to increase the energy at the expense of the 
luminosity, where the limit is set by the gradient and the available power.
The second is to increase the accelerating gradient and/or adiabatically extend the 
active linac by adding extra accelerating structure. This scheme was successfully adopted
at {\sc Lep} in raising the collision energy by more than a factor of two. Here the 
limit is set by the site length, the achieved gradient and the RF power.  Four 
super-conducting nine-cell cavity prototypes have been conditioned at 35~MV/m. 
This gradient 
would allow to push the {\sc Tesla} $\sqrt{s}$ energy up to 800~GeV, with a luminosity
of $5.8 \times 10^{34}$~cm$^{-2}$s$^{-1}$. The {\sc NLC} project aims at 
achieving 1~TeV, with luminosity of 3.4$\times 10^{34}$cm$^{-2}$s$^{-1}$, by doubling 
the number of components and, possibly, increasing the gradient as well.

Beyond these energies, the extensions of the SC and X-band technology are more 
speculative. In order to attain collisions at energies in excess of 1~TeV, with 
high luminosity, significantly higher gradients are necessary. Also, the number of 
active elements in the linac must be kept low enough to ensure reliable operation.
The two-beam acceleration scheme, presently being developed within the {\sc Clic} 
study~\cite{clic} 
at CERN, suggests a unique opportunity to extend the physics at $e^+e^-$ colliders to 
constituent energies of the order of the LHC energy frontier, and beyond. The recent 
test of a 30~GHz tungsten iris structure, successfully reaching an accelerating 
gradient of 150~MV/m~\cite{ctf}, for a 16~ns pulse, is highly encouraging for the 
continuation of these studies.

\section{The Neutral Higgs Boson Profile} 

Outlining the Higgs boson profile, through the determination of its mass, width, 
quantum numbers, couplings to gauge bosons and fermions and the reconstruction of the 
Higgs potential, stands as a greatly challenging programme. 
The spin, parity and charge-conjugation quantum numbers $J^{PC}$ of Higgs
bosons can be determined at the LC in a model-independent way. Already the 
observation of either $\gamma \gamma \rightarrow H$ production or 
$H \rightarrow \gamma\gamma$ decay sets $J \ne 1$ and $C=+$.
The angular dependence $\frac{d \sigma_{ZH}}{d \theta} \propto \sin^2 \theta$ and 
the rise of the Higgs-strahlung cross section $\sigma_{ZH} \propto~\beta 
\sim~\sqrt{s-(M_H+M_Z)^2}$ allows to determine $J^P = 0^+$ and distinguish the SM Higgs 
from a $CP$-odd $0^{-+}$ state $A^0$, or a $CP$-violating mixture of the two.
But the LC has a unique potential for verifying that the Higgs boson does its job of 
providing gauge bosons, quarks and leptons with their masses. This can be obtained by 
testing the relation $g_{HXX} \propto {M_X}$ between Yukawa couplings, 
$g_{HXX}$, and the corresponding particle masses, $M_X$, precisely enough. 
It is important to 
ensure that the LC sensitivity extends over a wide range of Higgs boson masses and that 
a significant accuracy is achieved for all particle species. Here, the LC adds the 
precision which establishes the key elements of the Higgs mechanism, as discussed in 
this section and summarised in Table~1.

\subsection{Couplings to Gauge Bosons}

\begin{figure}
\begin{center}
\begin{tabular}{c c}
\epsfig{file=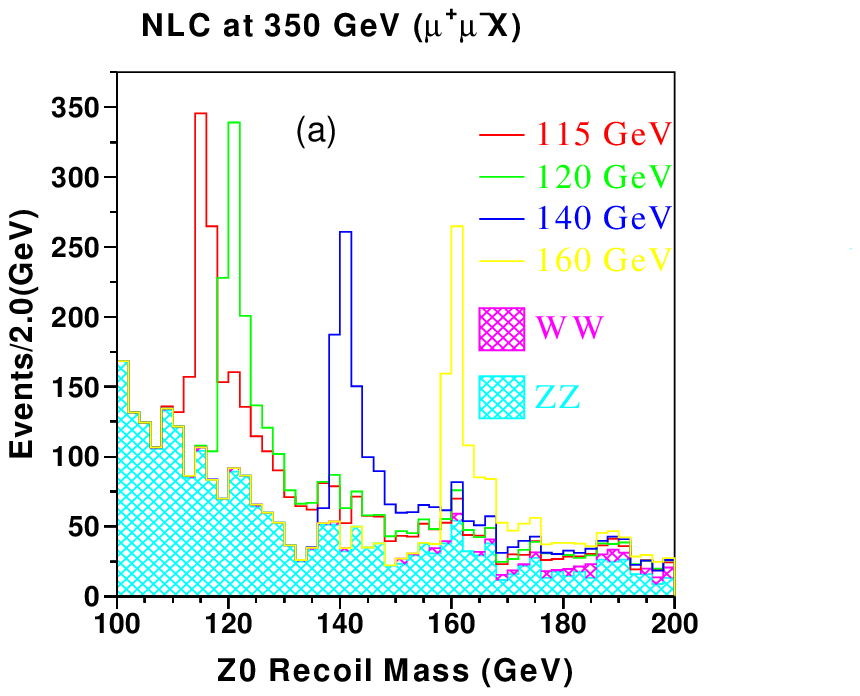,width=6.5cm,height=5.0cm} &
\epsfig{file=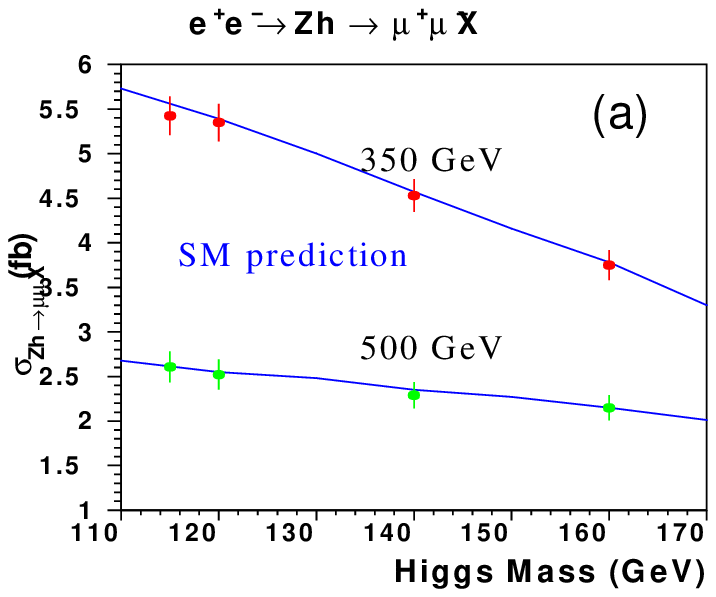,width=6.5cm,height=5.0cm} \\
\end{tabular}
\end{center}
\caption[]{\sl Left: Di-lepton recoil mass distributions with the Higgs boson signals 
for different mass values. Backgrounds are superimposed. Right: evolution of the 
Higgs-strahlung cross section vs. $M_H$ at $\sqrt{s}$=350~GeV and 500~GeV. The points 
with error bars represent the estimated uncertainties (from~\cite{Abe:2001wn}).} 
\label{fig:hz}
\end{figure}

The determination of the cross-section for the $e^+e^- \to H^0Z^0$ Higgs-strahlung 
process measures the Higgs coupling to the $Z^0$ boson and it is also a key input to 
extract absolute branching fractions from the experimental determination of products of 
production cross sections and decay rates.  At the LC, the use of the di-lepton recoil 
mass from the $Z^0 \rightarrow \ell^+ \ell^-$ decay (see Figure~\ref{fig:hz}) provides a
model-independent method which critically depends on the momentum resolution. An high 
precision central tracker and the addition of the beam-spot constraint guarantees 
$\Delta p/p \le 5 \times 10^{-5} p$~(GeV/c), which is $\simeq$12~times better than 
that obtained with the {\sc Lep} detectors. 
At the nominal luminosity expected at $\sqrt{s}$=350~GeV, 
of the order of 24000 Higgs bosons would be observable in one year of operation 
($=10^7$~s), of which 4000 in a model-independent way, if $M_H$=120~GeV. 
A relative accuracy on the $e^+e^- \to HZ$ cross section of 2.4-3.0\% can be achieved 
with 0.5~ab$^{-1}$ of data at $\sqrt{s}$ = 350~GeV, assuming 
120~GeV$< M_H <$160~GeV~\cite{hzxs}.

Since the recoil mass analysis is independent on the decay mode, 
it is also sensitive to non-standard
decay modes, such as $H \rightarrow {\mathrm{invisible}}$. Several theoretical models 
introduce an invisible $H$ decay width (possibly SUSY decays $\chi^0 \chi^0$, but also 
signatures of Radion-Higgs mixing or so-called Stealth models). Such invisible 
Higgs decays may be problematic at hadron colliders but are detectable both indirectly,  
by subtracting from the total decay width the sum of visible decay modes, and directly, 
by analysing the system recoiling against the $Z^0$ in the $e^+e^- \rightarrow HZ$ 
process, at the LC. The direct method generally provides with a higher accuracy, 
corresponding to a determination of BR($H \rightarrow {\mathrm{invisible}}$) to better 
than 5\%, so long as the invisible yield exceeds 10\% of the Higgs decay 
width~\cite{invisible}. 

The $WW$-fusion reaction, $e^+e^- \rightarrow WW \nu \bar \nu \to H \nu \bar \nu$,  
measures the $H^0$ coupling to the $W^{\pm}$ boson. A template analysis is based 
on $b$-tagged hadronic events at $\sqrt{s}$ = 350~GeV with large missing mass and 
missing energy. The main background is due to the Higgs-strahlung process, when 
$Z^0 \to \nu \bar{\nu}$. It is possible to extract $\sigma_{H \nu\nu}$ from a $\chi^2$ 
fit to the missing mass distribution, which efficiently discriminates between the 
two contributions~\cite{wwxs}. 
Overlapping accelerator-induced $\gamma \gamma \to {\mathrm{hadrons}}$
background can be suppressed by an impact parameter analysis of the forward produced 
particles, if the tracking resolution is small compared to the bunch 
length~\cite{Battaglia:1999ux}.
A relative accuracy of 2.6\% is obtained for $M_H$=120~GeV, which becomes 10\%, for 
$M_H$=150~GeV. However, since the branching fraction for the 
$H^0 \to W^*W$ decay increases sharply in this mass interval, the accuracy on the Higgs 
coupling to $W$ boson can be extracted with a small and constant uncertainties, when 
combining the results for production and decay processes, involving the same $HWW$ 
coupling.

\subsection{Couplings to Fermions}

Measuring the Higgs couplings to quarks and leptons precisely is one of the main aims 
of Higgs studies at the LC. The requirements of such analyses have driven the concept 
of the innermost vertex detector and the design of the interaction region. The issue 
here is to measure the charged particle trajectories accurately enough to distinguish the 
decays $H^0 \to b \bar{b}$ from $H^0 \to c \bar{c}$, and these from $H^0 \to gg$, by 
reconstructing the signature decay patterns of heavy flavour hadrons. Several 
independent studies have been performed which indicate that the BR($H^0 \to b \bar{b}$) 
can be measured to better than 3\%, BR($H^0 \to c \bar{c}$) to about 9-19\% and 
BR($H^0 \to g g$) to 6-10\%, if the Higgs boson is 
light~\cite{Battaglia:1999re,brientbr,Potter:2001ap}. The spread in the estimated 
accuracy for $c \bar{c}$ and $gg$ results is attributed in terms of the different 
simulation and data analyses methods adopted which are currently under study. 
It is interesting to observe that with
these experimental accuracies, the test of the coupling scaling with the fermion 
masses will be dominated by the present uncertainties on the latter.
The case of the top quark is of special interests as it is the only fermion with 
an ${\cal O}(1)$ Yukawa coupling to the SM Higgs boson. This coupling can be measured 
through a determination of the cross section 
$e^+e^- \to t \bar t H^0$~\cite{Dittmaier:1998dz}.

Tests of the mass generation mechanism in the lepton sector are also possible by 
studying the decays $H^0 \to \tau^+ \tau^-$ and $\mu^+ \mu^-$. The $\tau$ Yukawa 
coupling can be measured to 2.5-5.0\% using $\tau$ identification based on multiplicity 
and kinematics for 120~GeV$< M_H <$140~GeV~\cite{Battaglia:1999re,brientbr}. 
More recently, it has been shown that also the rare decay $H^0 \rightarrow \mu\mu$ is
observable at TeV-class LC and quite accurately measurable at a multi-TeV 
LC~\cite{Battaglia:2001vf}.  At 3~TeV, the relative accuracy on the muon 
Yukawa coupling is 3.5-10\% for 120~GeV$< M_H <$150~GeV. This would allow to test the
$g_{H\mu\mu}/g_{H\tau\tau}$ coupling ratio to a 5-8\% accuracy at a multi-TeV LC, which  
must be compared to the 3-4\% accuracy expected from combining the Muon Collider and 
TeV-class LC data for 120$<M_H<$140~GeV.

\subsection{Higgs Potential}

\begin{figure}
\begin{center}
\begin{tabular}{c c}
\epsfig{file=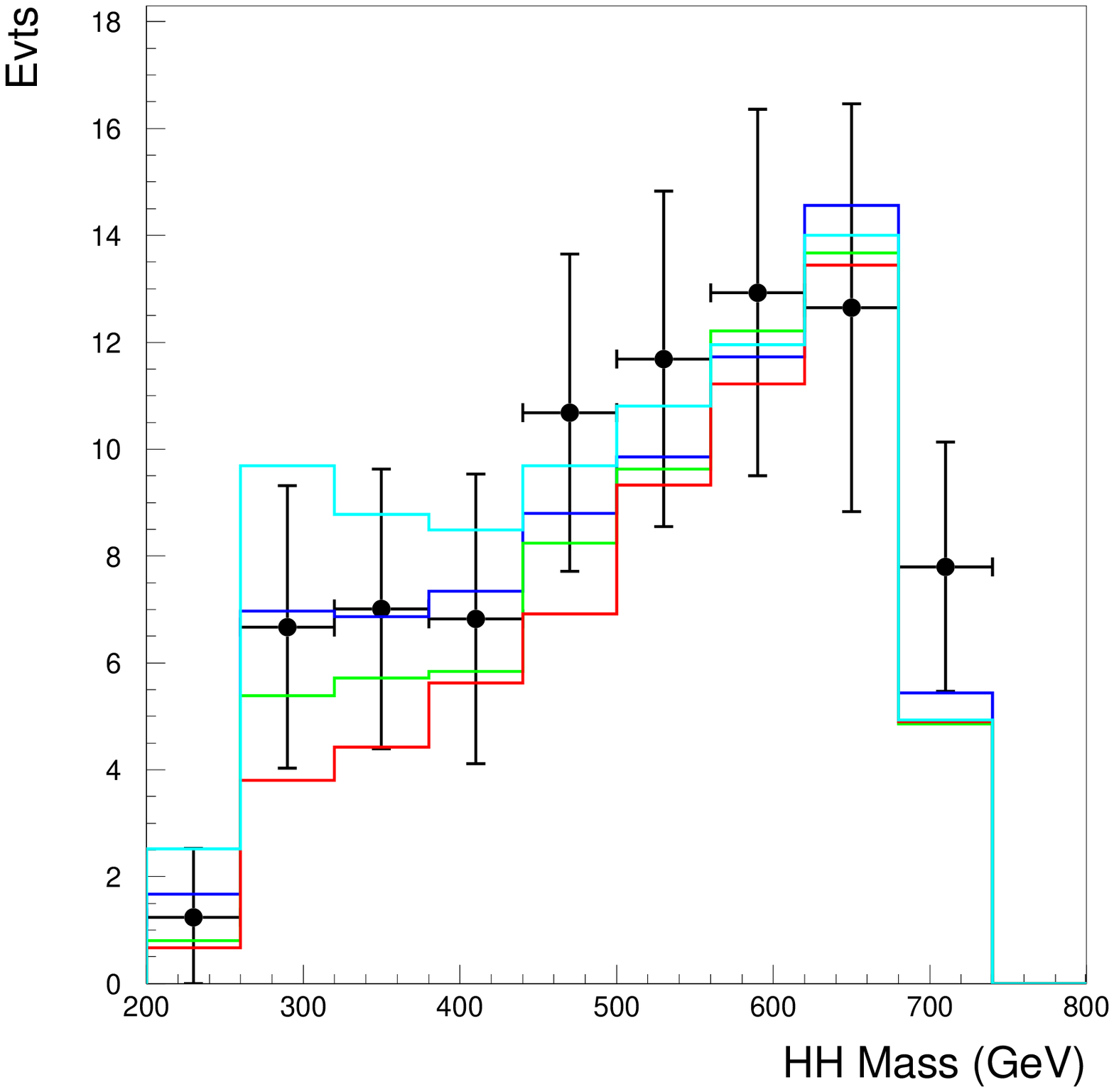,height=6.0cm,width=7.75cm,clip} &
\epsfig{file=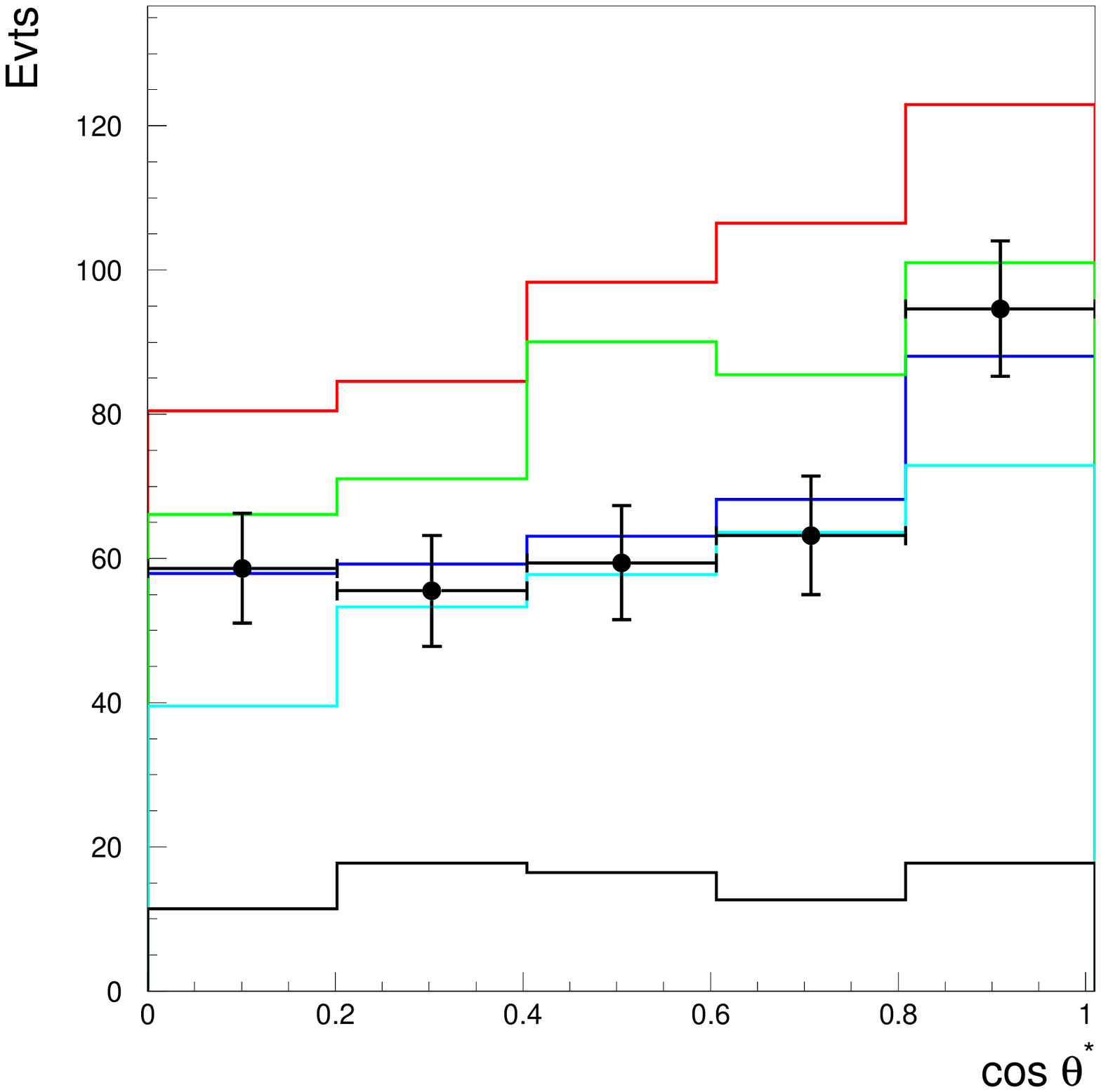,height=6.0cm,width=7.75cm,clip} 
\end{tabular}
\end{center}
\caption[]{\sl
Left: $HH$ invariant mass distribution for $HHZ$ events.
Right: reconstructed $|\cos \theta^*|$ distribution for $HH\nu\bar{\nu}$ events.
The lines give the expectations for $g_{HHH}/g_{HHH}^{SM}=$1.25,1.0,0.75 and 0.5. 
The points with error bars represent 1~ab$^{-1}$ of SM data at $\sqrt{s}$=0.8~TeV 
and 5~ab$^{-1}$ of SM data at $\sqrt{s}$=3.0~TeV respectively 
(from~\cite{Battaglia:2001nn}).}
\label{fig:hhh}
\end{figure}

A most distinctive feature of the Higgs mechanism is the shape of the Higgs potential, 
$V(\Phi^* \Phi) = \lambda (\Phi^*\Phi - \frac{1}{2}v^2)^2$. 
In the SM, the triple Higgs coupling, $g_{HHH}$, is related to the Higgs mass, $M_H$, 
through the relation $g_{HHH} = \frac{3}{2} \frac{M_H^2}{v}$, where $v$=246~GeV.
By determining $g_{HHH}$, in the double Higgs production processes 
$e^+e^- \to H H Z$ and $e^+e^- \to H H \nu\nu$~\cite{Djouadi:1999gv}, the above relation 
can be tested at the LC.
These measurements are made difficult by the tiny production cross sections and the 
dilution due to diagrams leading to double Higgs production, but not sensitive to the 
triple Higgs vertex. A LC operating at $\sqrt{s}$ = 500~GeV can measure 
the $HHZ$ production cross section to about 15\% accuracy, if the Higgs boson mass is 
120~GeV, corresponding to a fractional accuracy of 23\% on 
$g_{HHH}$~\cite{Castanier:2001sf}. Improvements can be obtained both by performing the 
analysis at very high energy and by introducing observables sensitive to the presence 
of the triple Higgs vertex (see Figure~\ref{fig:hhh})~\cite{Battaglia:2001nn}. On the 
contrary, the quartic Higgs coupling remains elusive, due to the smallness of the 
relevant triple Higgs production cross sections.

\begin{table}
\caption[]{\sl Summary of the accuracies on the determination of the Higgs boson 
profile at the LC. Results are given for a 350-500~GeV LC with 
${\cal{L}}$=0.5~ab$^{-1}$. Further improvements expected from a 3~TeV LC are also 
shown for some of the measurements.}
\begin{center}
\begin{tabular}{l|c|c|}
         & $M_H$            & $\delta(X)/X$ \\
         & (GeV)            & LC-500~~~~ $|$ LC-3000 \\
         &                  & 0.5~ab$^{-1}$ $|$ 5~ab$^{-1}$ \\ 
 \hline \hline
  $M_H$      & 120-180     & (3-5) $\times 10^{-4}$ \\
  $\Gamma_{tot}$ & 120-140 & 0.04-0.06 \\ \hline
  $g_{HWW}$ & 120-160 & 0.01-0.03 \\ 
  $g_{HZZ}$ & 120-160 & 0.01-0.02 \\ \hline
  $g_{Htt}$ & 120-140 & 0.02-0.06 \\
  $g_{Hbb}$ & 120-160 & 0.01-0.03 \\
  $g_{Hcc}$ & 120-140 & 0.03-0.10 \\
  $g_{H\tau\tau}$ & 120-140  & 0.03-0.05 \\ 
  $g_{H\mu\mu}$ & 120-140  & 0.15 $|$ 0.04-0.06 \\ \hline
  {\cal CP} test & 120   & 0.03  \\
  $g_{HHH}$ & 120-180 & 0.20 - -- $|$ 0.07-0.09 \\ \hline 
\end{tabular}
\end{center}
\label{tab:summary}
\end{table}
\begin{figure}
\begin{center}
\begin{tabular}{c c}
\epsfig{file=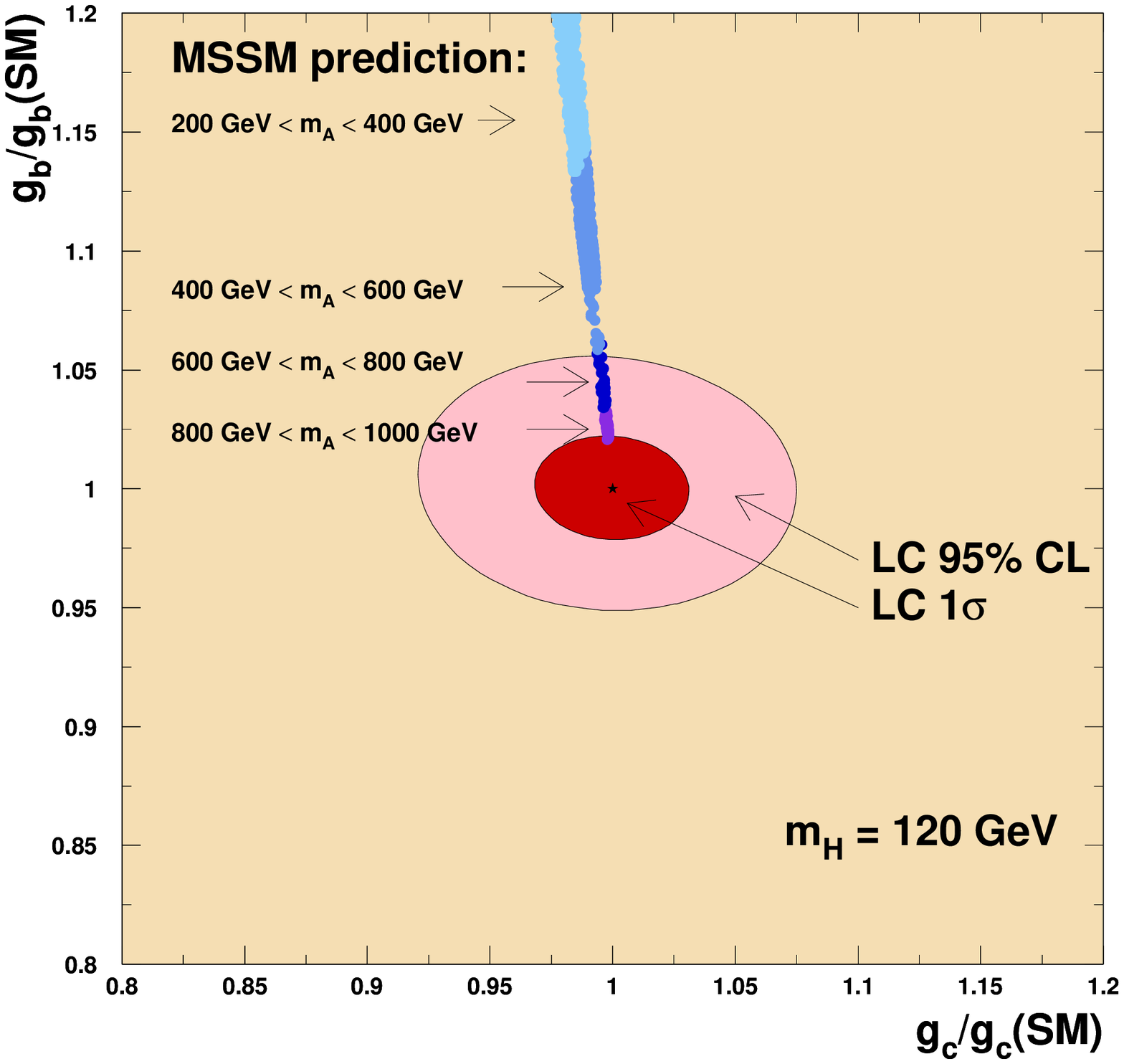,width=7.5cm,height=6.75cm,clip} &
\epsfig{file=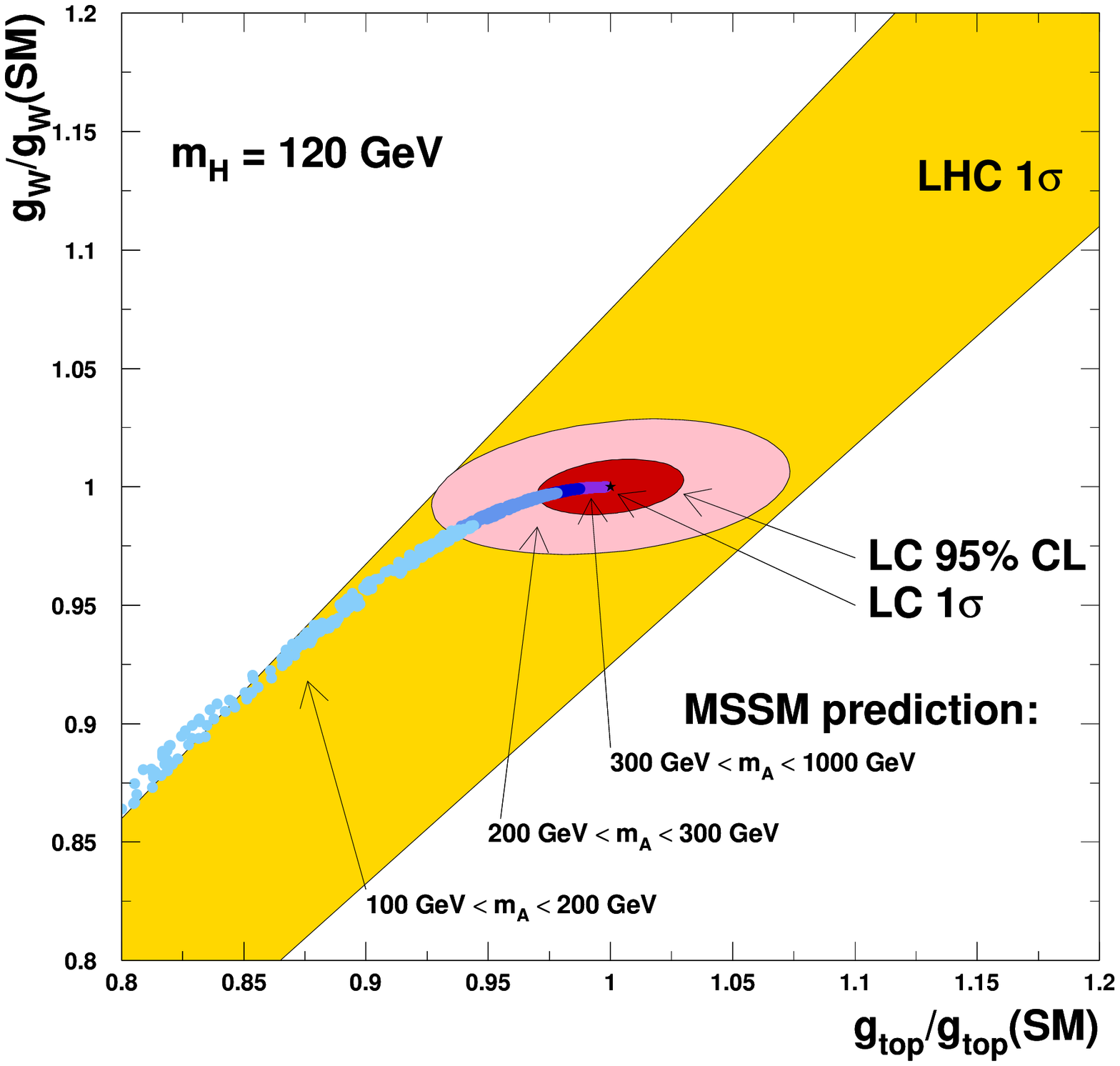,width=7.5cm,height=6.75cm,clip} \\
\end{tabular}
\end{center}
\caption[]{\sl Expected accuracies on the Higgs couplings from a global fit to the 
cross section and branching fraction determinations at the LC. Left: $g_{Hbb}$ vs. 
$g_{Hcc}$ and Right:  $g_{HWW}$ vs. $g_{Htt}$. Couplings are normalised to their 
SM predictions. The expected MSSM values are also shown for different values of 
$M_A$. Contours show the 1~$\sigma$ and the 95\% C.L. regions. The band on the right 
reproduces the expected sensitivity of the {\sc Lhc} data 
(from~\cite{Desch:2001xh}).}  
\label{fig:hfitter}
\end{figure}

\begin{figure}
\begin{center}
\epsfig{file=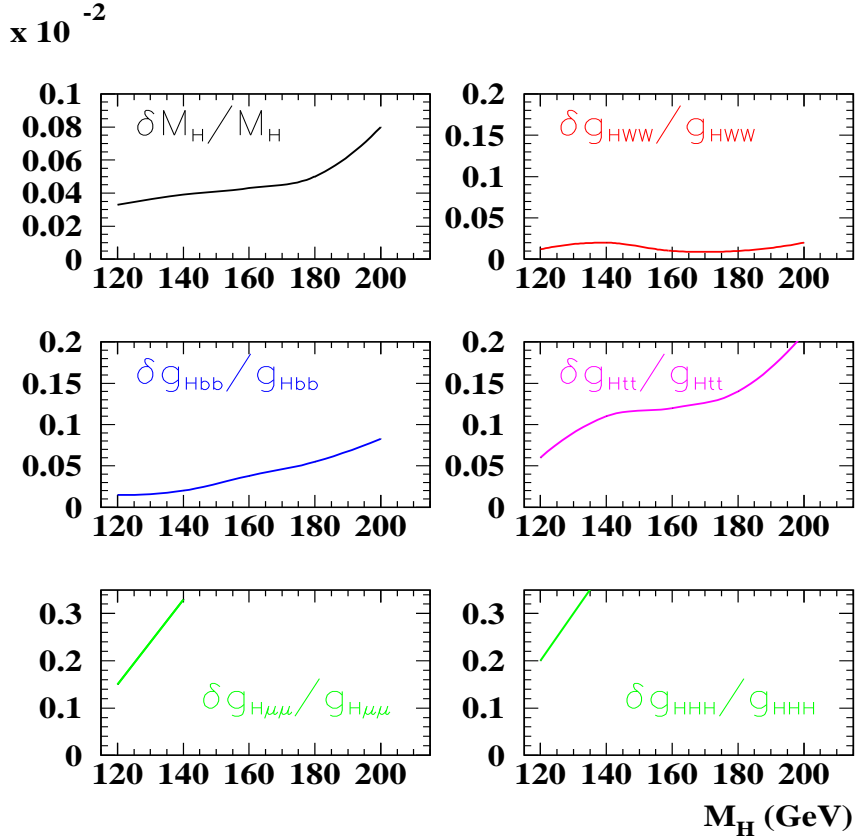,width=12.5cm,height=11.0cm,clip}
\end{center}
\caption[]{\sl Anticipated accuracies for the determination of the main 
Higgs properties at the LC, for $\sqrt{s}=350-800$~GeV, as a function of the 
Higgs boson mass.}
\label{fig:summary}
\end{figure}

\begin{table}
\caption[]{\sl Summary of the accuracies on the determination of a heavy Higgs boson 
profile at the LC. Results are given for a 500-800~GeV LC with 
${\cal{L}}$=0.5 and 1.0~ab$^{-1}$. Further improvements expected from a 3~TeV LC 
are also shown for some of the measurements.}
\begin{center}
\begin{tabular}{c|c|c|}
    & $M_H$ & $\delta X/X$\\
    & GeV   & LC-500/800 $|$ LC-3000\\
    &       & 0.5/1~ab$^{-1}$  $|$ 5~ab$^{-1}$ \\
\hline \hline 
$M_H$ & 240 & $9 \times 10^{-4}$ \\
$\Gamma_H$ & 240 & 0.12 \\ \hline
$\sigma(e^+e^- \rightarrow HZ)$ & 240 & 0.04 \\
BR($H \rightarrow ZZ$) & 240 & 0.10 \\
BR($H \rightarrow WW$) & 240 & 0.07 \\ \hline
BR($H \rightarrow b \bar b$) & 200 & 0.16 $|$ 0.04 \\
BR($H \rightarrow b \bar b$) & 220 & 0.27 $|$ 0.05 \\
\hline
\end{tabular}
\end{center}
\label{tab:heavy}
\end{table}
 
\subsection{What if the Higgs is heavier ?}

Precision electro-weak data indicate that a SM-like Higgs boson should be lighter than 
$\simeq$195~GeV. However, scenarios exists where the Higgs boson is heavier as a result 
of New Physics affecting the electro-weak observables. It is therefore important to 
assess the LC sensitivity to heavier bosons.
Analyses have considered the $HZ \rightarrow\ell^+\ell^-$, $q \bar q$ 
recoil mass at 500~GeV and the $H \nu \bar \nu$ process at 800~GeV. In order to extract 
$M_H$, $\Gamma_H$ and $\sigma$ a fit to the recoil mass spectrum can be performed while 
the $H \rightarrow WW$ and $ZZ$ branching fractions can be measured from the jet-jet 
mass in $HZ$~\cite{heavy}. The couplings to fermions are still accessible through 
$H \rightarrow b \bar b$ studied in $H \nu \bar \nu$. Results are summarised in 
Table~\ref{tab:heavy}.

\section{An Extended Higgs Sector}

Despite its successful tests, the Standard Model must be embedded into a more 
fundamental theory valid at higher energies, in order to explain its many parameters 
and cure remaining problems, such as the stabilisation of the Higgs potential. 
Supersymmetry is the simplest perturbative explanation for low-scale electroweak 
symmetry breaking, and, in its minimal realization (MSSM), predicts that 
the lightest Higgs boson must be lighter than about 130~GeV. In the MSSM, as in more 
general 2HDM extensions of the SM, the Higgs sector consists of two 
doublets, generating five physical Higgs states: $h^0$, $H^0$, $A^0$ and $H^{\pm}$.
The $h^0$ and $H^0$ states are CP even while the $A^0$ is CP odd. The masses of the 
CP-odd Higgs boson, $M_A$, and the ratio of the vacuum expectation values of the two 
doublets $\tan \beta = v_2/v_1$ are free parameters.
\begin{figure}[h!]
\begin{center}
\begin{tabular}{c c}
\epsfig{file=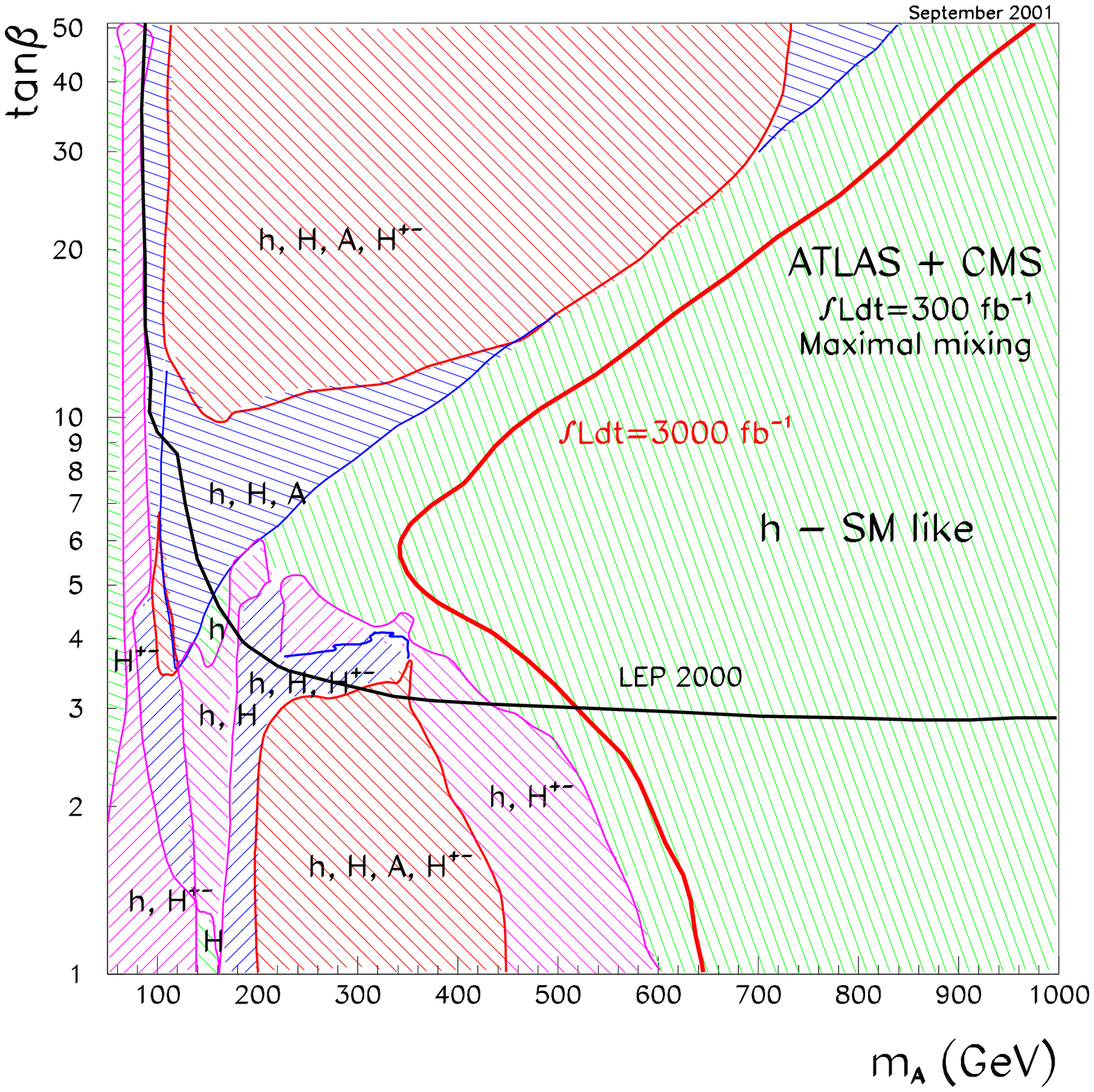,width=8.5cm,height=8.0cm,clip} &
\epsfig{file=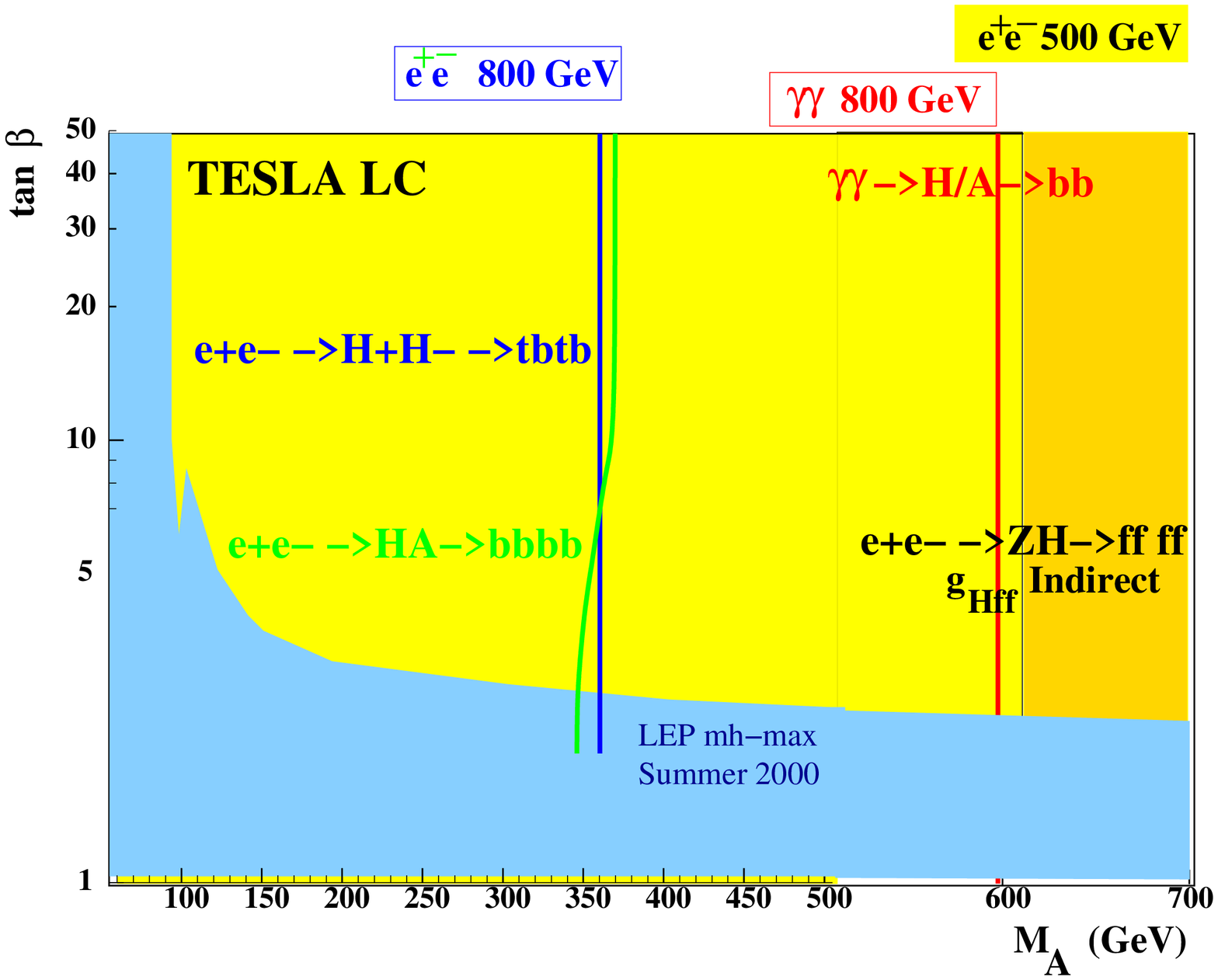,width=6.5cm,height=8.75cm,clip} \\
\end{tabular}
\vspace*{-0.5cm}
\end{center}
\caption[]{\sl Reach for MSSM Higgs boson discovery in the $M_A$ - $\tan \beta$ 
plane at {\sc Lhc} and LC. 
Left: expected combined performance of {\sc Atlas} and {\sc Cms} with 
${\cal{L}}$=300~fb$^{-1}$. The wedge at large $M_A$ and moderate $\tan \beta$ values 
corresponds to the parameter region where only the lightest 
Higgs is observable. Right: expected LC performance at $\sqrt{s}$=500 and 800~GeV. 
The direct search extends up to $M_A \simeq$~350~GeV, while the indirect sensitivity 
extends to $M_A$ values of 600-700~GeV. A $\gamma \gamma$ collider with 
$\sqrt{s_{ee}}$=800~GeV should match this limit with the direct observation of single 
$H^0$ and $A^0$ production.}
\label{fig:matb}
\end{figure} 
The study of the lightest neutral MSSM Higgs boson, $h^0$, follows closely that
of the SM $H$ discussed above and those results remain in general valid. 
The mass and coupling patterns of the other bosons vary with the model parameters. 
However, in the decoupling limit the $H^{\pm}$, $H^0$ and $A^0$ bosons are expected 
to be heavy and to decay predominantly into quarks of the third generation.
Establishing their existence and determining their masses and
main decay modes, through their pair production $e^+e^- \to H^0 A^0$ and $H^+H^-$
would represent a decisive step in the understanding of the Higgs sector, which will 
most probably require extended LC operations at $\sqrt{s} \ge$~500~GeV. 

\subsection{Indirect Sensitivity}

The precision study of the Higgs couplings to fermions and gauge bosons may already 
reveal its SM or Supersymmetric nature, before the direct observation of heavier 
bosons. In fact, in the SM we expect the ratios of 
Yukawa couplings to be proportional to those of the particle masses.
On the contrary, in Supersymmetry, couplings to up-like and down-like fermions are 
shifted w.r.t. their SM predictions as
$\frac{BR(h \rightarrow f_u \bar{f_u})}{BR(h \rightarrow f_d \bar{f_d})}
\propto \frac{1}{tan^2 \alpha \tan^2 \beta} \simeq 
\frac{(M^2_h - M^2_A)^2}{(M^2_Z + M^2_A)^2}$. This can be exploited to distinguish the 
SM $H^0$ from the Supersymmetric $h^0$ with precise determination of its 
couplings (see Figure~\ref{fig:hfitter}).
If they would result to be incompatible with those predicted by the SM, informations 
can be extracted on the value of the $A^0$ mass, $M_A$, which is a fundamental parameter
in Supersymmetry. Several analyses of the indirect sensitivity to the MSSM provided by 
the accurate determination of the Higgs couplings have been conducted as a function of 
$M_A$. Results show consistently that for $M_A<$650~GeV, the MSSM $h^0$ can be 
distinguished from the SM $H^0$, mostly from its $b \bar{b}$, $c \bar{c}$ and $WW$ 
couplings (see Figure~\ref{fig:matb}). Furthermore, SUSY sbottom-gluino and 
stop-higgsino loops may shift the effective $b$-quark mass in the $hbb$ couplings: 
$\Delta m_b \propto \mu ~M_{\tilde{g}}~\tan \beta~f(M_{\tilde{b_1}}, M_{\tilde{b_2}}, 
M_{\tilde{g}})$. This becomes important in specific regions of the parameter space, 
at large $\tan \beta$ and small $M_A$ values, where the $b \bar{b}$ coupling 
can be dramatically suppressed. These effects can be accurately surveyed at the LC.

\subsection{Direct Sensitivity}

If the heavy Higgs bosons are above pair-production threshold, the 
$e^+e^- \to H^0$, $A^0 \to b \bar b$, $e^+e^- \to H^+ \rightarrow t \bar b$ 
or $H^+ \to W^+ h^0$, $h^0 \to b \bar b$ processes will provide with very distinctive, 
yet challenging, multi-jet final states with multiple $b$-quark jets, which must  
be efficiently identified and reconstructed. Example analyses have shown 
that an accuracy of about 0.3\% on the boson masses and of $\simeq 10\%$ on the 
product $\sigma \times BR$ can be obtained at the LC~\cite{hpm,ha}. 

In addition, the $\gamma \gamma \to A^0$ and $H^0$ process at the $\gamma \gamma$ 
collider, is characterised by a sizable cross section which may probe the heavier 
part of the Higgs spectrum, beyond the $e^+e^-$ reach. Finally, a scan of the $A^0$ 
and $H^0$ thresholds at the $\gamma \gamma$ collider can in principle resolve a 
moderate $A^0 - H^0$ mass splitting~\cite{ggha,Muhlleitner:2001kw}, 
which could not otherwise be observed at other colliders. 

\subsection{CP violation}

Extensions of the SM may introduce new sources of CP violation, through additional 
physical phases whose effects can be searched for in the Higgs sector.
Supersymmetric one-loop contributions can lead to differences in the decay rates of 
$H^+ \to t \bar{b}$ and $H^- \to \bar{t} b$, in the MSSM with complex 
parameters~\cite{Christova:2002ke}. This CP asymmetry is expressed as 
$\delta CP = \frac{\Gamma(H^- \rightarrow b\bar{t})-\Gamma(H^+ \rightarrow t\bar{b})}
{\Gamma(H^- \rightarrow b\bar{t})+\Gamma(H^+ \rightarrow t\bar{b})}$ and it can amount 
to up to $\simeq 15$\%. As the leading contributions come from loops with $\tilde{t}$, 
$\tilde{b}$ and $\tilde{g}$, $\delta^{CP}$ is sensitive to these parameters.
With the expected statistics of $e^+e^- \to H^+H^- \to t \bar{b} \bar{t} b$ at 
$\sqrt{s}$=3~TeV and assuming realistic charge tagging performances, a 3~$\sigma$ effect
for ${\cal{L}}$=5~ab$^{-1}$ would be observed for an asymmetry $|\delta^{CP}|$=0.10.

\subsection{NMSSM}

While the MSSM has been the main model for surveying the SUSY Higgs phenomenology 
so far, Supersymmetry may be realised in a non-minimal scenario. The introduction of an 
additional Higgs singlet has been proposed as a natural explanation 
of the value of the $\mu$ term in the MSSM Higgs potential. The resulting NMSSM Higgs 
sector has seven physical Higgs bosons and six free parameters.
\begin{figure}[h!]
\begin{center}
\begin{tabular}{c c}
\epsfig{file=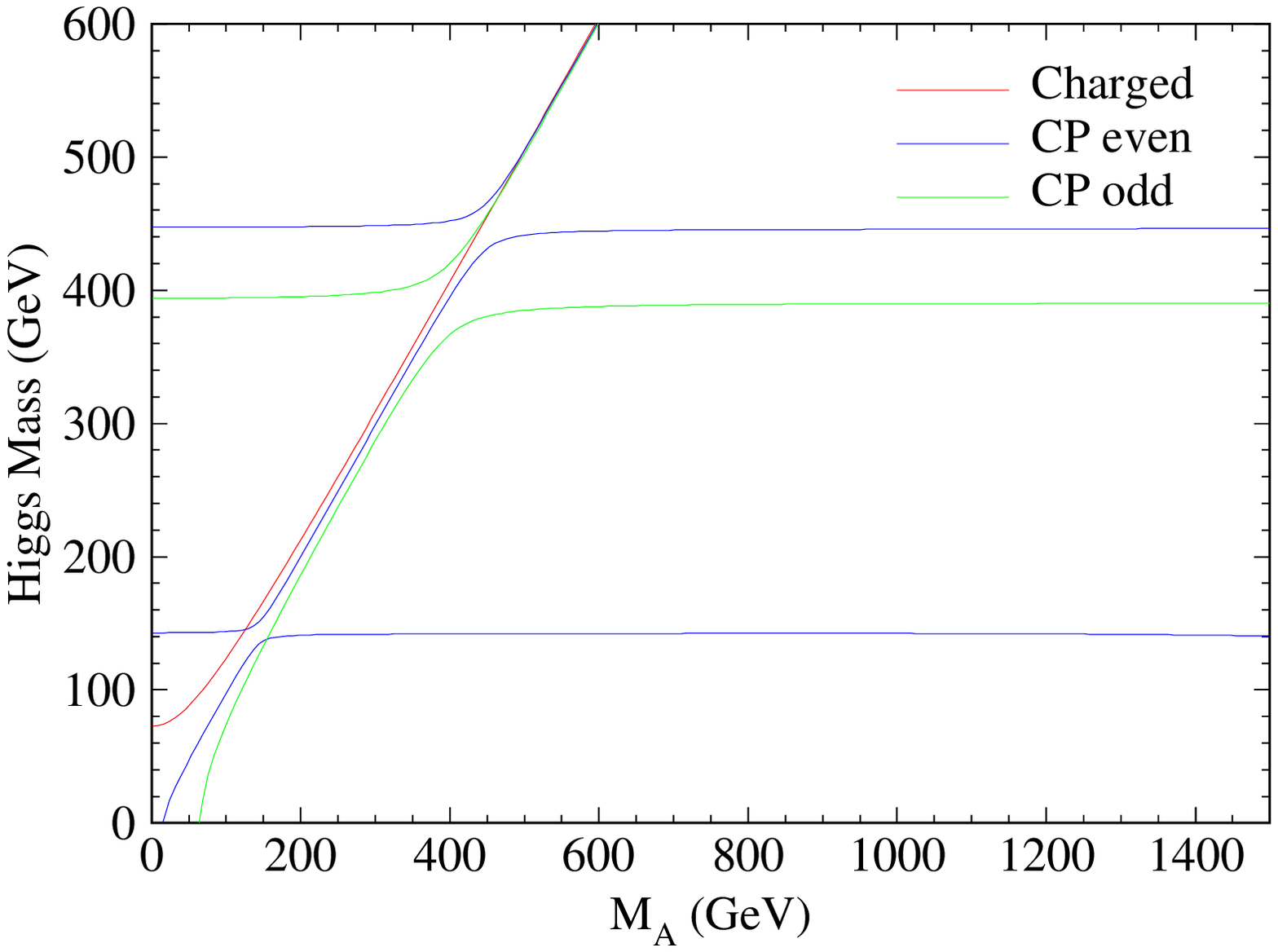,width=7.5cm} &
\epsfig{file=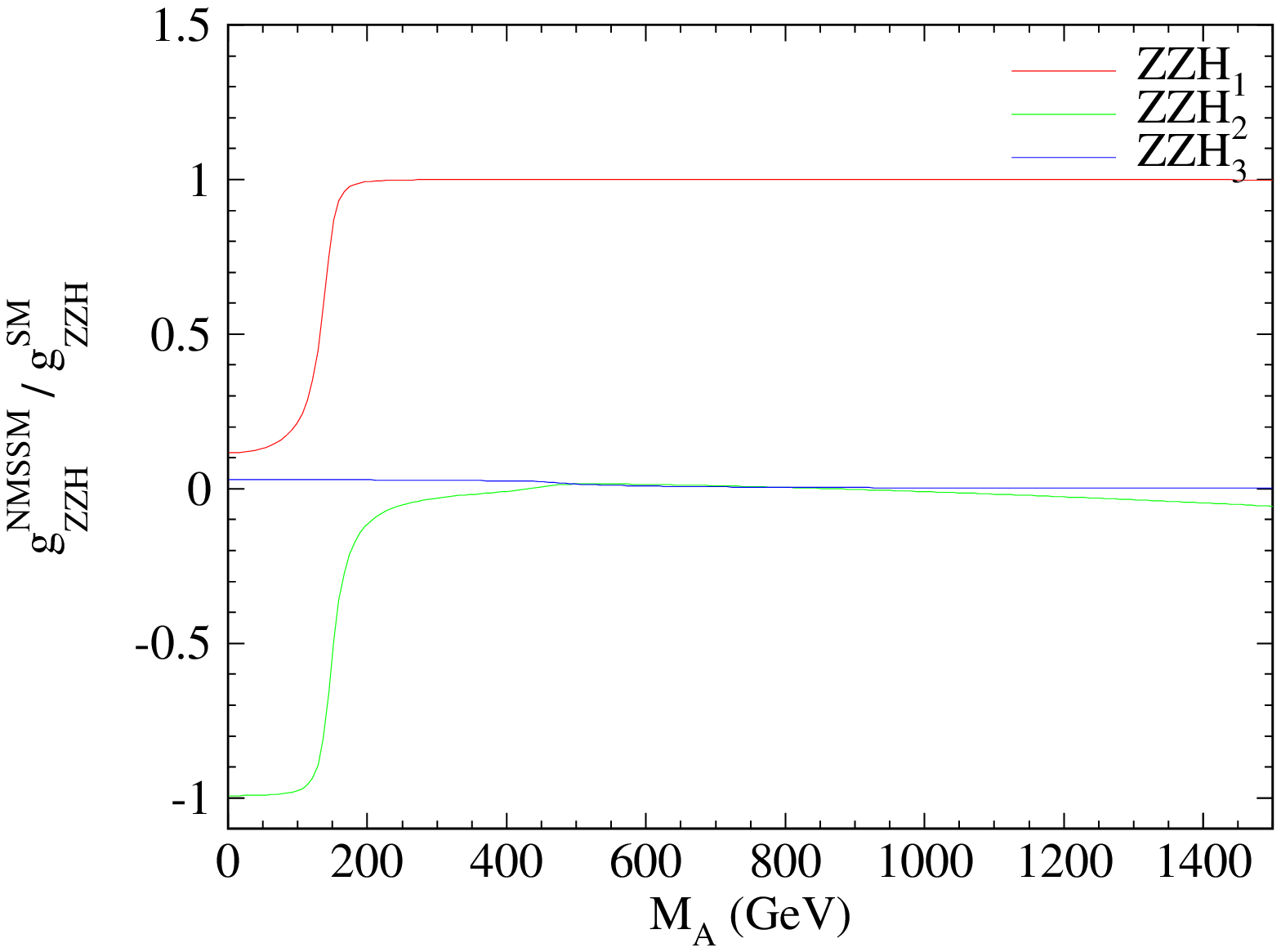,width=7.5cm} \\
\end{tabular}
\end{center}
\caption[]{\sl NMSSM Higgs masses (left) and couplings to $Z^0$ bosons normalised to 
SM value (right) as a function of $M_S$, for a typical set of model 
parameters (from~\cite{miller}).}
\label{fig:nmssm}
\end{figure}
It has been shown that the prospects for Higgs discovery at the {\sc Lhc} are not 
undermined in this scenario, provided that the full integrated luminosity of 
${\cal{L}}$=300~fb$^{-1}$ is considered~\cite{Ellwanger:2001iw}.
This would lead to an interesting phenomenology with two scalar Higgs bosons possibly 
within reach of a TeV-class LC, one light pseudo-scalar and four heavy bosons almost 
degenerate in mass (see Figure~\ref{fig:nmssm})~\cite{miller}.

\section{The Higgs Boson and the Radion\\ in scenarios with Extra-Dimensions}

\begin{figure}[h!]
\begin{center}
\begin{tabular}{c c}
\epsfig{file=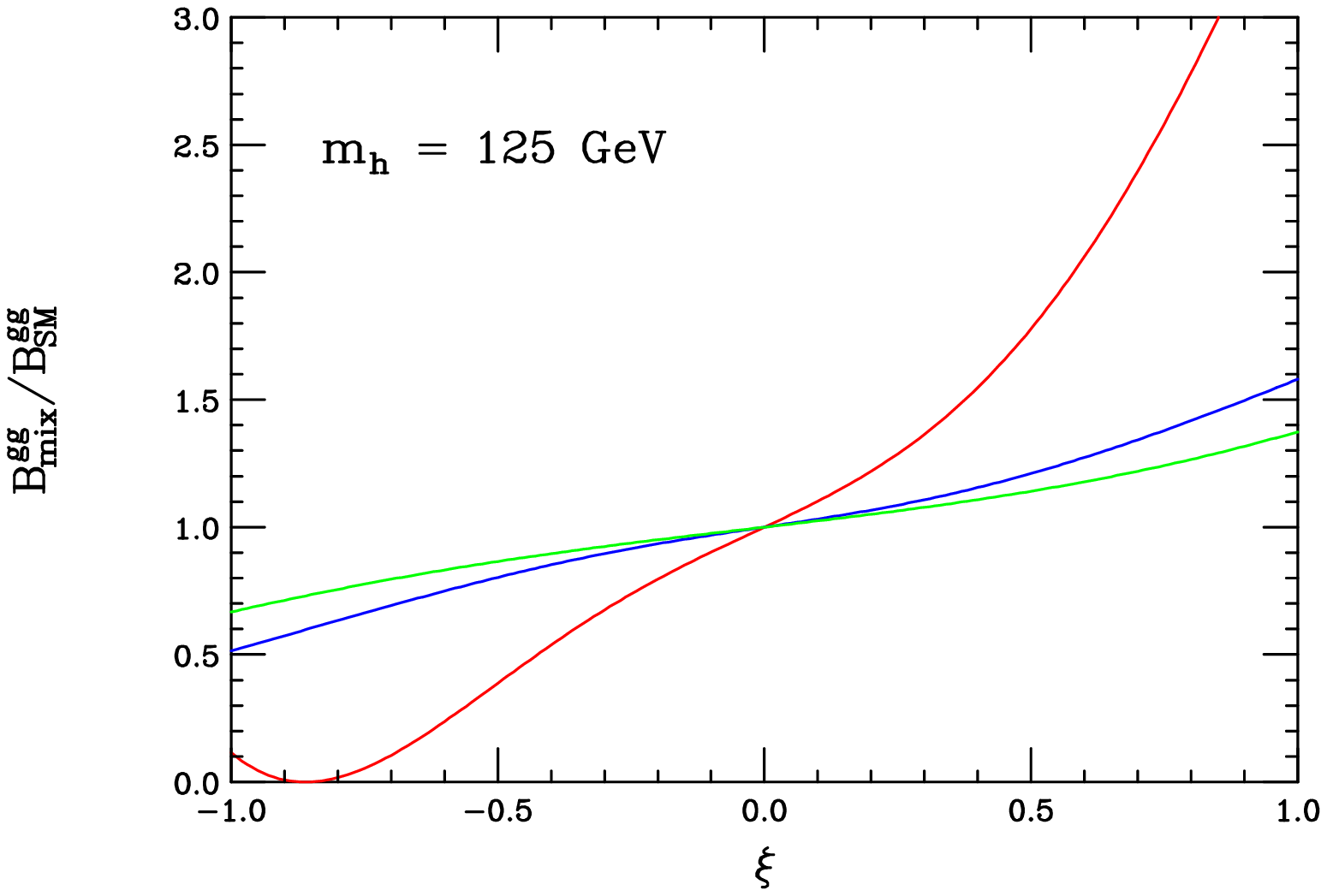,width=7.5cm,height=6.5cm} &
\epsfig{file=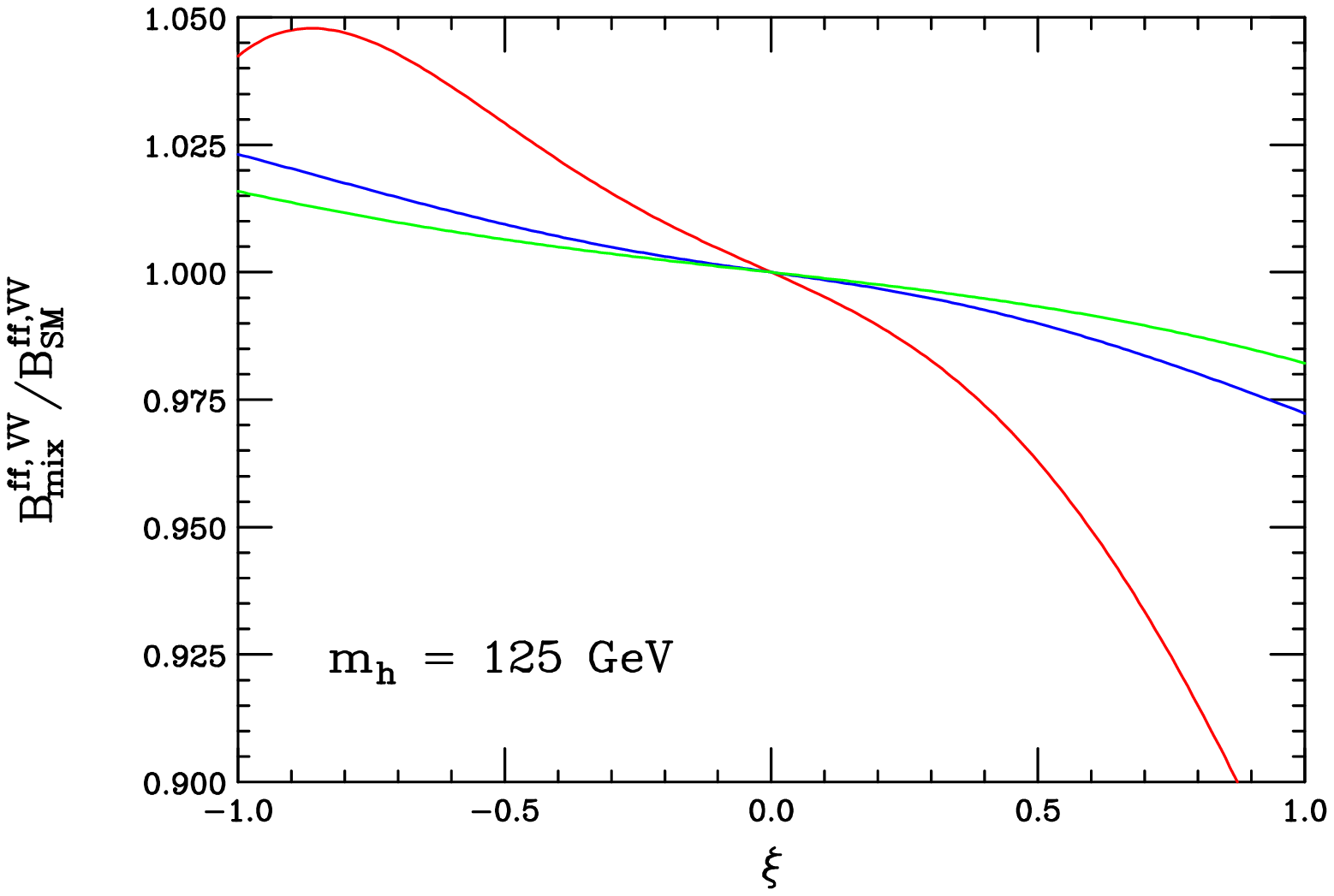,width=7.5cm,height=6.5cm} \\
\end{tabular}
\end{center}
\caption[]{\sl Modified Higgs couplings to gluons (left) and fermions and vector 
gauge bosons (right) induced by Radion-Higgs mixing. The effects are shown, as function 
of the mixing parameter, $\xi$, for three sets of values of the Radion mass and the 
$\Lambda$ scale of the theory (from~\cite{Hewett:2002nk}).}  
\label{fig:radion}
\end{figure}

The hierarchy problem, originating from the mismatch between the electroweak scale, 
defined by the Higgs field vacuum expectation value $v$ = 246~GeV and the Planck scale, 
has motivated the introduction of models with hidden extra dimensions. 
In the original construction of the Randall-Sundrum formulation~\cite{Randall:1999vf}, 
the SM particles live on a brane, while gravity expands on a second, parallel brane and 
in the bulk. This scenario introduces a new particle, the Radion, which represents the 
quantum excitation of the brane separation. By mixing with the Higgs field, the Radion 
modifies the Higgs couplings to SM particles and thus its decay branching fractions 
BR($H \rightarrow f \bar{f}$)~\cite{Hewett:2002nk,Dominici:2002jv}. Examples are shown 
in Figure~\ref{fig:radion} for $M_H$=125~GeV. The typical accuracies for the branching 
fraction determination at the LC make these shifts significant enough to reveal the 
radion effects over a large fraction of the parameter space, including regions where 
evidence of the radions and of the extra-dimensions could not be obtained directly.

\section{Conclusions}

The present theory of fundamental interactions and the experimental evidence 
strongly suggest that the electro-weak symmetry is broken via the Higgs mechanisms, 
while particles acquire their mass through their interaction with the scalar Higgs 
field. The Higgs boson is heavier than 114~GeV and possibly lighter than about 210~GeV. 
Within this scenario we expect the {\sc Lhc} to discover such Higgs boson and a high 
energy, high luminosity $e^+e^-$ linear collider to accurately determine its properties.
These precisions can be obtained within a realistic run plan scenario. Charting the 
Higgs profile will not only clarify if the Higgs mechanism is indeed responsible for 
electro-weak symmetry breaking and mass generation. It will also elucidate 
the nature of the Higgs particle. As the discovery of Hispaniola deceived Columbus on 
the essence of his achievement, the Higgs sector may have more and different properties 
than those expected in the SM or even considered in this paper. Higgs physics at the LC 
may thus provide the first clues on New Physics beyond the SM and possibly a new world 
of particles, as predicted in Supersymmetry. 
While most of the mapping of the lighter Higgs profile can be optimally 
performed at a LC with centre-of-mass energies in the range 300-500~GeV, there 
are measurements which will require higher energies. An upgrade of the centre-of-mass 
energy to $\simeq$~1~TeV and a second stage multi-TeV LC should complete the measurement
of the Higgs properties with the needed accuracy and extend the sensitivity to heavier 
Higgs bosons, beyond the {\sc Lhc} reach, depending on the nature of the Higgs sector. 
In all cases, linear colliders add crucial information to previous data and to the data
that the {\sc Lhc} will obtain.

\vspace{0.5cm}

{\sl It is a pleasure to thank the organisers for their invitation and for a most 
pleasant Conference. I am grateful to Albert~De~Roeck, Klaus~Desch, Daniele~Dominici, 
Jack~Gunion, Joanne~Hewett, Tom~Rizzo, Ian~Wilson and Peter~Zerwas for suggestions and 
discussion.}

\end{document}